\documentclass[aps,amsfonts,prx,twocolumn,superscriptaddress,showpacs]{revtex4-1}

\usepackage[normalem]{ulem}
\usepackage{float}
\usepackage[usenames]{color}
\usepackage{graphicx}
\usepackage{ulem}
\usepackage{amsfonts}
\usepackage{amsmath}
\usepackage{natbib}

\usepackage{caption}
\usepackage{subcaption}
\begin{document}

\title{Leader-cell-driven epithelial sheet fingering}
\author{Yanjun~Yang}
\affiliation{Department of Applied Physics and Center for Theoretical Biological Physics, \\Rice University, Houston TX, 77251-1892}
\author{~Herbert~Levine}
\affiliation{Department of Bioengineering and Center for Theoretical Biological Physics, \\Rice University, Houston TX, 77251-1892}
\affiliation{Departments of Physics and Bioengineering, \\Northeastern University, Boston, MA 02115}
\date{\today}

\begin{abstract}
Collective cell migration is crucial in many biological processes such as wound healing, tissue morphogenesis, and tumor progression. The leading front of a collective migrating epithelial cell layer often destabilizes into multicellular finger-like protrusions, each of which is guided by a leader cell at the fingertip. Here, we develop a subcellular-element-based model of this fingering instability, which incorporates leader cells and other related properties of a monolayer of epithelial cells. Our model recovers multiple aspects of the dynamics, especially the traction force patterns and velocity fields, observed in experiments on MDCK cells. Our model predicts the necessity of the leader cell and its minimal functions for the formation and maintenance of a stable finger pattern. Meanwhile, our model allows for an analysis of the role of supra-cellular actin cable on the leading front, predicting that while this observed structure helps maintain the shape of the finger, it is not required in order to form a finger. In addition, we also study the driving instability in the context of continuum active fluid model, which justifies some of our assumptions in the computational approach. In particular, we show that in our model no finger protrusions would emerge in a phenotypically homogenous active fluid and hence the role of the leader cell and its followers are often critical.  
\end{abstract}
 
\maketitle
\section{Introduction}
Collective cell migration drives many crucial physiological processes including wound healing, tissue morphogenesis and tumor progression \cite{SilberzanReview17, MartinScience97, MartinJCellSci09, LecuitNRMCB07}. Previous experimental studies  have investigated how a group of epithelial cells move coordinately to close a wound both in vivo and in vitro \cite{BementARCDB11Rev, LadouxCOCB16Rev}. Pioneering measurements using convenient in vitro model systems such as Madin-Darby canine kidney (MDCK) cells  have focused on the mapping of mechanical properties of these 2D expanding epithelial sheets \cite{SilberzanNCB14, TrepatNatPhys09, TrepatNatPhys12, TrepatNatPhys14, LadouxPNAS05, SilberzanBPVF10, SilberzanBPOP11}.  These efforts have uncovered details of the dynamics including the traction force patterns and velocity fields in collectively migrating cellular sheets. This observed behavior arises via a complex mechanical and biochemical set of processes which involve various mechanisms at different scales \cite{SilberzanNCB14, SilberzanPNAS07, SilberzanBiophys14}.

Many of these experiments observe multicellular fingering-like protrusions, aka fingers, on the leading front of spreading tissues (Fig \ref{fig2V}B).  These experiments also find that fingers often associate with ``leader" cells on the front boundary \cite{SilberzanPNAS07, SilberzanNCB14, SilberzanBPVF10, SilberzanBPOP11}.  Leader cells are also seen when tissues expand in three dimensional extracellular matrix \cite{ewald, MarcusNC17}. A leader cell has a noticeably different morphology than cells further behind in the finger (the ``followers") or ones in the tissue bulk. For example, the leader cell is often much larger than a regular cell in size (Fig \ref{fig2V}B) \cite{SilberzanPNAS07, SilberzanBPVF10} and it exerts a larger traction force. From a biological perspective, a leader is a specialized phenotype \cite{SilberzanPNAS07, SilberzanNCB14}, expressing a different complement of proteins reflecting a different state of the genetic network; this has been directly established in  the biological literature \cite{MarcusJCS19, WongNC15}.  In addition,  intermediate morphologies between the leader cell and regular cells are observed for cells we refer to as followers. This graded behavior presumably arises via reciprocal coupling to the nearby leader cell which induces a partial leader phenotype. For example, the cell size gets larger closer to the fingertip \cite{SilberzanPNAS07, SilberzanBPVF10}. However, the necessity of and the effects caused by the leader cell as far as finger formation is concerned is still being debated \cite{SilberzanBioPMod10, SilberzanNCB14, BasanPNAS13, ZimmermannEP14, LeePRE17, AlertPRL19, SavinRSOS18, SpatzNC18}. In addition to the leader cell, supracellular actomyosin cables are often observed on the sides of fingers \cite{SilberzanPNAS07}.  It is well-established that these cables can be crucial in wound healing, especially for epithelial closure on a non-adhering substrate; experimental and theoretical studies show that a localized wound cannot close on a non-adhering surface without a supracellular actomyosin cable creating an effective purse-string contraction around the wound \cite{LadouxNC15, LadouxNM14, YangSM18}. The role of the actomyosin cable in the protruding finger requires further understanding.

Possible mechanisms underlying a fingering instability of a planar propagating interface have been investigated by mathematical modeling. Both particle-based and continuous models have proposed to describe this instability \cite{SilberzanBioPMod10, NGovIntBio15, HakimComBio13, BasanPNAS13, BiNatPhys15, BiPRX16}. These models have made it clear that a variety of mechanisms can cause such an instability, including the role of growth behind the advancing front and the effects of polarization of the tissue by cell-cell interactions. Thus leader cells are not necessary for this aspect of fingering. However, it is well-established in the theory of interfacial pattern formation \cite{kkl85} that stable fingers require additional mechanisms (such as crystalline anisotropy for solidification) as the default is repeated tip-splitting. Leader cells are almost always observed at the tips of stable fingers \cite{BasanPNAS13, SavinRSOS18, SilberzanPNAS07, SilberzanNCB14, SilberzanBPVF10}. and early work that incorporated leader cells in a variety of ways were the only ones that circumvented this problem \cite{SilberzanBioPMod10, NGovIntBio15}. However, it is clearly important to note that most of these earlier works did not attempt to make contact with detailed biophysical data regarding the velocity and (especially) traction force patterns that accompany the finger morphology.  Among the above works, the continuous models are based on active media theory, following the seminal work of Toner and Tu  \cite{TuPRL95, TuPRE98, TuAP05, MarchettiRMP13} on a hydrodynamical approach to systems of self-propelled particles. However, these models have not considered the possible dynamics and the function of leader cells and in addition have not actually generated stable nonlinear finger states, being content to predict unstable fronts. 

In this article, we develop a subcellular-element-based computational model to address these unsolved issues. The computational approach is developed as an extension of our recent work focused on explaining the mechanics of expanding monolayer sheets of MDCK cells \cite{YangSM18, ZimmermannPNAS16, BasanPNAS13}. We extend our previously presented framework to include a model of a leader cell as a special cell with phenotypically altered parameters, including a larger self-propelled force, stronger adhesion, and the ability to actively attract nearby follower cells. Our model also takes into account the intermediate follower phenotype that occurs between the leader cell and the cells very far away. This intermediate phenotype partially adopts leader cell behavior and has a graded behavior from the fingertip to the base. The supracellular actomyosin cable on sides of the finger is also included in this framework \cite{YangSM18}. As we will see, the model successfully accounts semi-quantitatively for observed traction force patterns and cellular dynamics; this does not occur in our model without the leader and follower cell phenotypes. Thus, we predict that taking into account the leader cell and the graded follower behavior are often necessary in order to form a finger with observed biophysical behavior. In addition, we also predict that the supracellular cable on sides of the finger is not necessary for its formation, though the cable contributes to the traction force and to the finger morphology. In parallel, we partially justify these findings via the study of an active fluid model. Starting from the basic Toner-Tu equations, we incorporate a curvature-based force on the leading interface so as to simulate the extra pulling force from a leader cell; this is analogous to the idea presented in \cite{SilberzanBioPMod10}. We should also mention that a curvature dependence of cell traction was proposed in Ref \cite{Rausch}, albeit within the context of a static tissue constrained by designed micro-patterned surfaces. We show that in our formulation the interface is stable without the curvature-based force. Instead, only when the force at the leading front is strong enough and transmittable can the fingering protrusions emerge. In other words, both the leader and the graded behavior are required. All told, our study shows that finger patterns can be explained by models that couple biomechanical effects with phenotypic variation and hence this type of approach will often be necessary to get an accurate picture of some critical aspects of collective cell motility.

\section{Key mechanisms in the computational model} 
\subsection{Basic model}
\begin{figure*}[htbp] 
\centering
    \includegraphics[width=.8\linewidth]{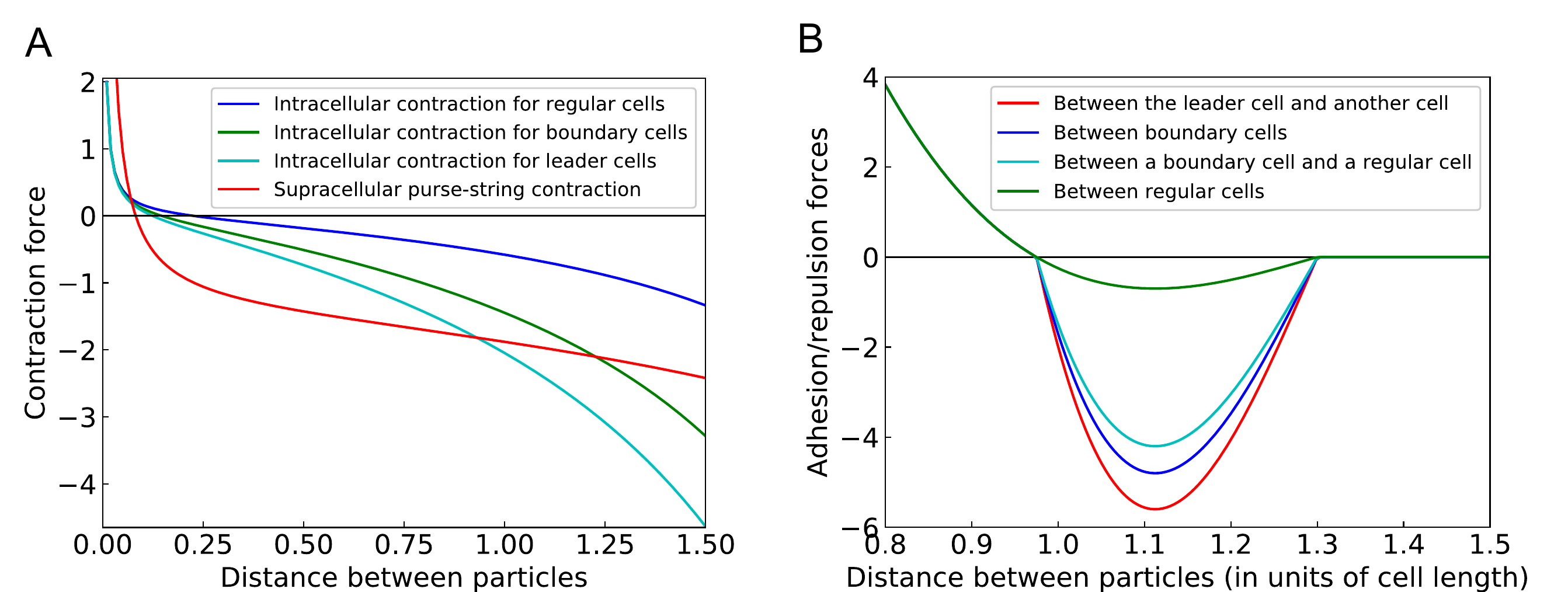}     
  \caption{Forces between sub-cellular elements. A. Contraction force. B. Intercellular adhesion/repulsion force.}
  \label{fig1} 
\end{figure*}
Here we present our subcellular-element-based model for leader-cell driven finger formation. In general, a variety of methods can be used to describe the collective motility of epithelial sheets as a collection of interacting dynamical cell-type objects. These range from simple agent-based models in which the cell is treated as a point particle \cite{viscek, levine} to more complex formulations that include morphology degrees of freedom either on a lattice \cite{Graner1992, Szabo2010, Albert2016} or in the continuum \cite{Shao2010, Shao2012, Ziebert2012, BiNatPhys15, BiPRX16}. Here we choose to use a sub-cellular element approach \cite{Sandersius2008, Sandersius2011, BasanPNAS13, ZimmermannPNAS16, YangSM18}  in which a cell is represented by two interacting point-like objects. This approach is the simplest  that allows for the prediction of traction force patterns. This type of model has successfully explained the mechanical state of expanding tissue \cite{TrepatNatPhys09} as well as the interaction of expanding tissue with surface inhomogeneities \cite{trepat-inhomogeniety}. It is important to note that this class of model does not aim to determine directly from the data the value of actual biophysical constants relevant for the complex processes underlying cell motility; instead our focus is in connecting mechanisms to patterns in the data, as will become clear below.

In detail, we implement a two-dimensional sub-cellular element approach in our simulation model. Each cell is represented by two sub-cellular elements, the front and the rear element. Each element is sell-propelled with a self-propulsion force $\vec{m}$. This self-propulsion force is regulated by contact inhibition of locomotion (CIL), meaning the force is suppressed by cell-cell contact and it aligns away from a cell's neighbors (more details in SI and Ref \cite{ZimmermannPNAS16}). The front and the rear element interacts with each other through an intracellular contractile force $\vec{f}_{contr}$ (Fig \ref{fig1}A). Elements from different cells interact via an intercellular force $\vec{f}_{rep/adh}$. This force is repulsive at short distances, attractive at longer distances, and becomes zero further away, thereby taking into account volume exclusion and cell-cell adhesion (Fig \ref{fig1}B). We consider the cell to be moving on a stiff substrate and assume a uniform friction coefficient $\xi$ between each sub-cellular element and the substrate. The velocity of each sub-cellular element is given by force balance,
\begin{equation}
\vec{v} = \frac{1}{\xi} (\vec{m} + \vec{f}_{contr} + \vec{f}_{rep/adh}).
\end{equation}
The position of each element is updated by a simple Euler scheme $\Delta \vec{x} = \vec{v} dt$. The net traction force exerted on each element by the substrate is $\vec{m} - \xi \vec{v}$. 

\subsection{Generalized Voronoi construction}
The finger patterns we focus on in this paper arise in epithelial sheets that maintain local confluency (i.e. have no visible gaps between cells). Our computational strategy does not directly encode the boundary of a cell and hence we have adopted a modified version of the Voronoi construction to construct a representation of the actual cellular structure contained within our simulated tissue. Specifically, all points $x$ within the plane are classified as follows. We compute the distance from that test point to each of the subcellular particle locations for each of our computational cells; these are labeled as $d_{i,1} (x)$, $d_{i,2} (x) $. From these we determine the effective separation as $D_i = d_{i,1} +d_{i,2} - S_i$ where $S_i$ is the separation between the two elements of cell $i$. This definition guarantees that the separation is exactly zero for any point lying on the line segment between the element locations, We also introduce a maximum separation $D_{max}$. Our algorithm is then
that point $x$ is inside cell $i$ if $D_i$ is the smallest among all the cells and it is below $D_{max}$; otherwise the point is outside the tissue. This is a generalization to elliptical structures of a method developed in Ref \cite{teomy}.

\subsection{Leader cell}
To model the observed phenotypic variation occurring at the leading edge, we introduce the concept of a leader cell. These cells are endowed with properties gleaned from experimental observations. We do not model the processes of cell-cell communication that lead to the emergence of leader cells and instead we adopt a phenomenological strategy in which leader cells are selected randomly among cells on the leading front. Once being selected as a leader cell, both sub-elements in the cell are given a larger self-propulsion force $m$, which is regulated by CIL in a same manner as for regular cells (Detailed parameters are in SI). To balance the increased the self-propulsion force, the intracellular contraction $f_{contr}$ and the friction coefficient $\xi$ are also increased. Also, the leader cell has a larger maximum cell-cell adhesion with neighboring cells. 

The leader cell attracts follower cells within a certain range $R$, endowing them with an enhanced self-propulsion force $\delta m$. This assumption is based on the aforementioned observations regarding follower cell size and also on data regarding the difference between follower and bulk cells with respect to the Notch signaling pathway \cite{WongNC15}. Also, there is direct evidence \cite{MarcusNC17} regarding leader cells in 3d that they secrete diffusible chemical signals such as VEGF which modulate follower cell motility. This additional self-propulsion force points towards the leader cell and has a magnitude based on its distance to the leader cell, decreasing as the distance increases (more details in SI). Meanwhile, the maximum cell-cell adhesion between follower cells within the same range $R$ is also increased in a similar fashion. The cell-substrate friction coefficient and cell division threshold length (introduced in the subsection below) are increased in this fashion as well (more details in SI). These mechanisms reflect graded behavior from the leader cell to the follower cells to the bulk cells. We will show this behavior is necessary to form a persistent finger; it is not enough to put in a single isolated leader. The algorithm checks the eligibility of the cell to stay a leader by comparing its position to that of the tip of the finger every step; a leader cell switches back to a regular cell if it falls behind the fingertip. A new leader cell might then emerge at the fingertip (more details in SI). This emergence of new leaders has also been seen experimentally \cite{reinhart-king}.

\subsection{Supracellular cable}
Neighboring epithelial cells can connect their actin cortices across their respective membranes to form a cable. Here we follow our previous strategy that was shown capable of reproducing force patterns during the healing of small circular wounds \cite {YangSM18}. To model this supracellular actomyosin cable on sides of the fingers, we label these cells as ``boundary cells". Elements of ``boundary cells" are mechanically linked with their nearest neighbors via an extra contractile force $f_{cable}$ similar to the intracellular contraction $f_{contr}$ (Fig \ref{fig1}B). These links of neighboring cells on the sides of the fingers will then form a purse-string contraction cable, which models the supracellular actomyosin cable observed on sides of fingers. In addition to the cable, ``boundary cells" have a larger maximum cell-cell adhesion with each other and their neighboring regular cells (more details in SI and Ref \cite{YangSM18}). We will show this cable is necessary to maintain the smoothness of a finger and it plays an important role in giving rise to the observed traction force patterns. 

\subsection{Cell proliferation and other effects}
Cells can change their size due to mechanical effects in our simulation; we do not explicitly consider cell growth due to other factors. Our model assumes a cell division rule based on cell size. Each cell divides at a given rate when it exceeds a certain threshold length \cite{ZimmermannPNAS16, YangSM18}. At each simulation step, the cell length of each cell is checked, and if it is longer than the threshold length, the cell divides with a fixed probability. Upon division, a new sub-cellular element is inserted at a random position within a certain range of each element from the original cell, forming two new cells. Finally, we distinguish motile and non-motile cells in our model. For a motile cell, the front element has a larger magnitude of the self-propulsion force ($m_f$) than the rear element ($m_r$). Conversely, a non-motile cell has a same magnitude of the self-propulsion force for both the front and the rear elements. Leader cells are always motile. We adopt one aspect of the alignment of cellular motility used in our earlier work, namely each cell tends to align its motility force with its average velocity by switching between motile and non-motile states. Compared with CIL, this alignment contribution plays a relative minor role for the resultant tissue mechanics \cite{ZimmermannPNAS16}. 

\section{Results from the computational model}
\begin{figure*}[htbp]
\centering
    \includegraphics[width=.85\linewidth]{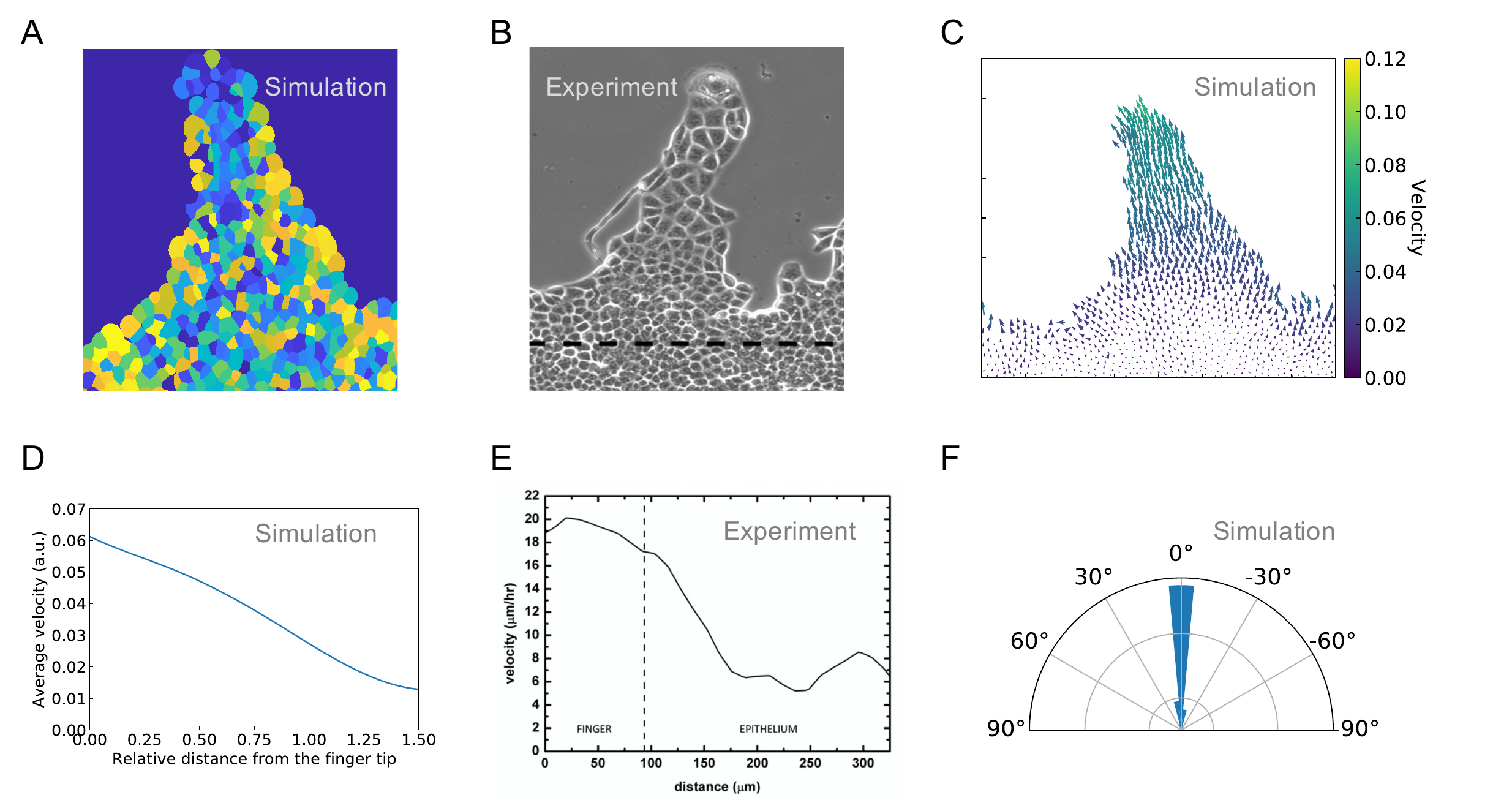}     
  \caption{Properties of an example finger: velocity. A. Voronoi cell representation of a finger from our simulation; details of our Voronoi construction are presented in the text. The coloring is just for visualization purposes.  B. Finger image from experiment \cite{SilberzanBPVF10}. C. Velocity profile. D. Average velocity for different distance from the fingertip. The velocity for each particle that has the same distance to the fingertip is averaged. Distance is normalized with the finger length, 0 is the tip, 1 is the base. Result from one finger is included in this subfigure. F. Average velocity from experiment \cite{SilberzanBPVF10}. F. Statistics for the velocity orientation of finger cells. The number of particles are summed up based on their velocity directions. The direction is defined by the relative angle to the orientation of the finger, i. e., $0^{\circ}$ means perfectly parallel to the finger orientation. Note: Figure  A, C, D and F are taken from the same simulation at time 610 in simulation units. The parameters are listed in Table S1 in SI. }
  \label{fig2V} 
\end{figure*}

\begin{figure*}[htbp] 
\centering
    \includegraphics[width=.85\linewidth]{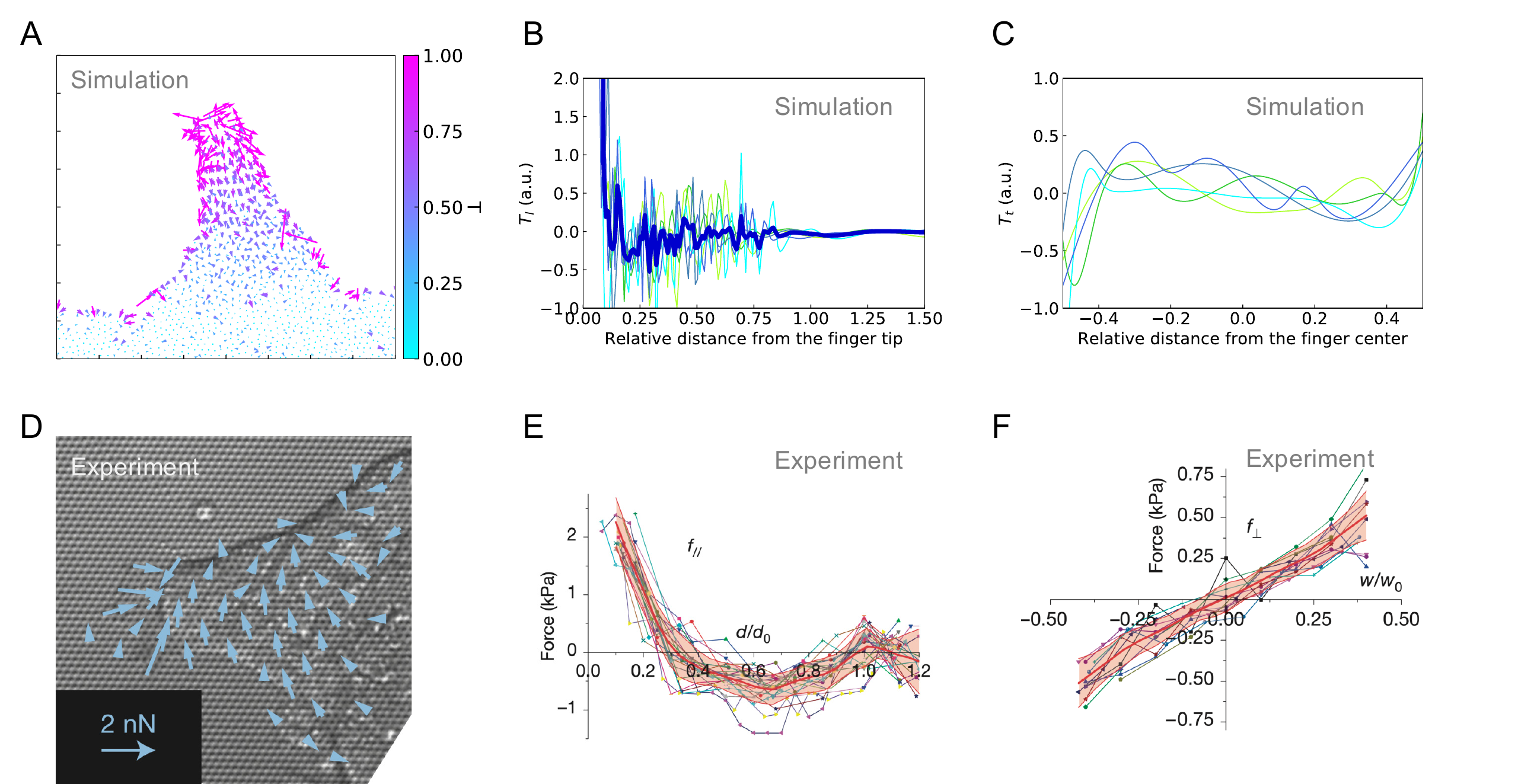}     
  \caption{Mechanical properties of an example finger: traction force. A. Traction force pattern. B. Average longitudinal traction force for different distance from the fingertip (The longitudinal component of the traction force for each particle has the same distance to the fingertip is averaged. Distance is normalized with the finger length, 0 is the tip, 1 is the base. Results from five different fingers are included in this subfigure. The thick blue line is the average of the results from these five fingers.). C. Average transverse traction force for different distance from the finger center (The transverse component of the traction force for each particle has the same distance to the finger center is averaged. Distance is normalized with finger width, -0.5 is the left boundary, 0 is the center, 0.5 is the right boundary. Results from five different fingers are included in this subfigure.). D. Traction force pattern from experiment \cite{SilberzanNCB14}. E. Average longitudinal traction force from experiment \cite{SilberzanNCB14}. F. Average transverse traction force from experiment \cite{SilberzanNCB14}.  Note: Figure  A-C are from the same simulation in Figure \ref{fig2V} at time 610 in simulation units. The parameters are listed in Table S1 in SI. }
  \label{fig2T} 
\end{figure*}

Our computational model can simulate the formation of fingers and obtain predictions for mechanical properties including traction force patterns. To demonstrate this, we initialize our simulation by seeding the cells in the center of a long rectangle box surrounded by hard walls. Cells will proliferate and eventually reach a steady state, forming a strip of cells. After that, we remove the wall on the upper side. Cells driven by CIL will move into the open space and leader cells will emerge at the front, based on our algorithm (SI, Movie S1). We observe that we can always get a finger-like protrusion if we have a persistent leader cell (Fig \ref{fig2V}A, SI Movie S1, which is similar to finger protrusions observed in experiment, Fig \ref{fig2V}B). In most cases, the leader cell has a tendency to move more-or-less upward. This arises because of the CIL effect, namely the leader is trying to move away from its followers.  The follower cells usually follow behind the leader cell and their combined dynamics gives rise to a prototypical velocity field, an example of which s shown in (Fig \ref{fig2V}C). The average velocity as a function of distance to the fingertip can be directly determined and we find that the largest velocity emerges at or near the fingertip and decreases from the fingertip to base (Fig \ref{fig2V}D). The velocity of finger cells are mostly directed along the finger (Fig \ref{fig2V}F), which matches the experimental observations. \cite{SilberzanBPVF10, SilberzanBPOP11}. For example, Fig \ref{fig2V}E presents the average velocity taken from experiment \cite{SilberzanBPVF10}. To compare with Fig \ref{fig2V}D, we need to notice that their finger region only contains what in our model is merely the fingertip. 

Our model can also provide us the traction force for each sub-cellular element at each time step (Fig \ref{fig2T}A, which is similar to experimental measurements, for example, Fig \ref{fig2T}D  \cite{SilberzanNCB14}. Note: Each cell is measured on two points in our simulation, while the traction force is usually measured per area in experiment.). Inside a finger, the traction force is highly dynamic. Locally, these force can be positive or negative (Fig \ref{fig2T}A). We calculate the average traction force in the parallel (longitudinal, $T_l$) and perpendicular (transverse, $T_t$) directions of the finger as a function of distance to the fingertip (Fig \ref{fig2T}B, \ref{fig2T}C). The longitudinal force $T_l$ at the fingertip is very large and positive (pointing from the fingertip to the base). After a few lines of cells, this large traction force becomes slightly negative and reaches the minimum near the middle of the finger (Fig \ref{fig2T}B). We can see this longitudinal traction force pattern is quite similar to the experimental result (Fig \ref{fig2T}E). The transverse force $T_t$ is largest at the side boundaries of the finger and points towards the center (Fig \ref{fig2T}C  \cite{SilberzanNCB14}). This inward force comes from the supracellular actomyosin cable and helps maintain the integrity of the finger. The transverse force pattern is also similar to the experiment (Fig \ref{fig2T}F \cite{SilberzanNCB14}). We do note that our transverse force is more concentrated on the boundary as compared to the more diffuse pattern seen in the experiment, There are several possibilities for this discrepancy, For example, our measurement is more refined and could capture the large force on the boundary while the experiment might have averaged this out. Perhaps our cable is too strong or perhaps the actual modification of the actin organization extends some distance in to the bulk from the edge, as opposed to what we have assumed. In any case, our traction data does a reasonable job of mimicking (and thereby explaining) the mechanisms underlying the data given the rather extreme simplification inherent in our cellular representation.

\begin{figure*}[htbp] 
\centering
    \includegraphics[width=.65\linewidth]{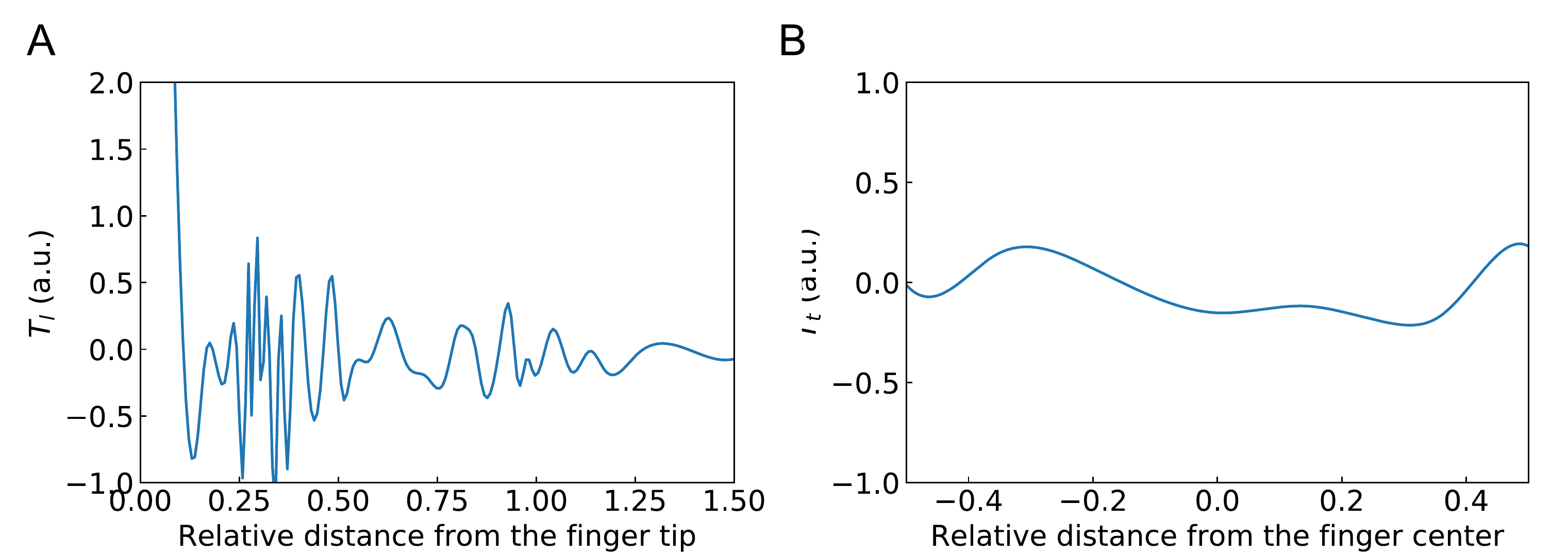}     
  \caption{Mechanical properties of a simulated finger without the supracellular cable on its sides. A. Average longitudinal traction force for different distance from the fingertip. The longitudinal component of the traction force for each particle has the same distance to the fingertip is averaged. Distance is normalized with the finger length, 0 is the tip, 1 is the base. Result from one finger is included in this subfigure.). B. Average transverse traction force for different distance from the finger center (The transverse component of the traction force for each particle has the same distance to the finger center is averaged. Distance is normalized with finger width, -0.5 is the left boundary, 0 is the center, 0.5 is the right boundary. Result from one finger is included in this subfigure.) Note: Both figures are from the same simulation at time 550 in simulation units. The parameters are listed in Table S1 in SI (parameters for the cable are void). }
  \label{fig3} 
\end{figure*}

\begin{figure}[htbp] 
\centering
    \includegraphics[width=1.0\linewidth]{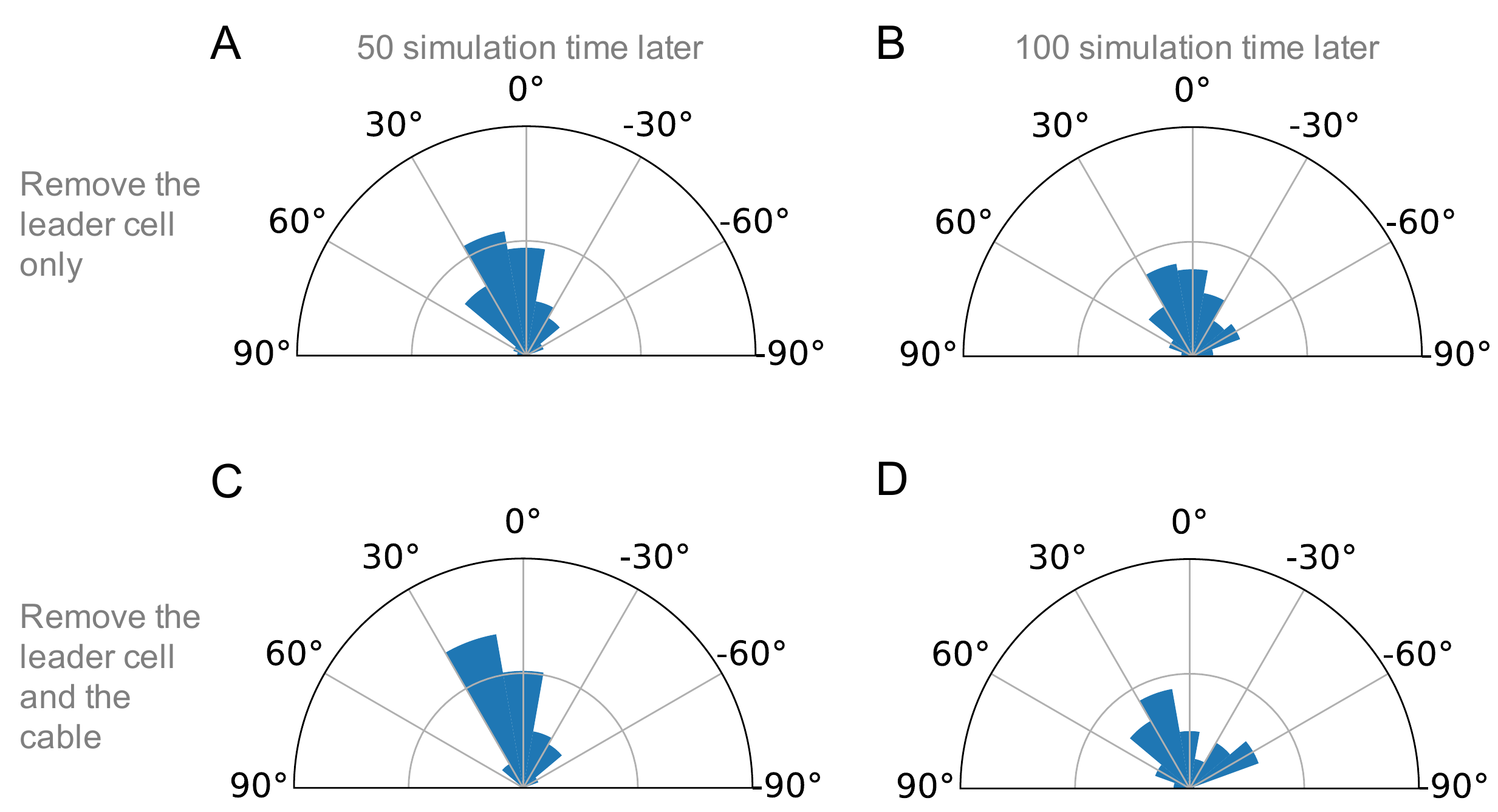}     
  \caption{ Statistics for the velocity orientation of finger cells. The number of particles are summed up based on their velocity directions. The direction is defined by the relative angle to the orientation of the finger, i. e., $0^{\circ}$ means perfectly parallel to the finger orientation. A, B. 50 and 100 simulation time after removing the leader cell only. C, D. 50 and 100 simulation time after removing the leader cell and the cable. }
  \label{fig4} 
\end{figure}

In absence of the leader cell, no finger can emerge (SI, Movie S2). In particular CIL alone is not a sufficient mechanism to form a finger. Instead, CIL stabilizes the entire leading front since it is uniform in the absence of any phenotypic perturbation. In other words, a leader is necessary in our model. The leader cell introduces an asymmetrical velocity field which is fastest on the fingertip and gradually decreases from the tip to the base (Fig \ref{fig2V}D, Ref \cite{SilberzanBPVF10}). In addition, the guidance of the leader cell prompts the follower cells to move along the finger, which enhances the polarity of cells within the finger (Fig \ref{fig2V}F, Ref \cite{SilberzanBPOP11}). If we remove the leader cell in an already existing finger, this polarity will vanish very quickly (Fig \ref{fig4}). Exactly what causes this type of coupling between leader and follower is not determined by our phenomenological approach. Possible mechanisms include either biochemical signaling or mechanical coupling (or a combination). This will be further discussed later on.

In absence of the cable on sides of the finger, a finger might emerge as long as there is a leader cell (SI, Movie S3). However, without the cable the predicted transverse traction force pattern would be very different from what is usually observed in experiments. In particular, the large inward pointing traction force on side borders of the finger will disappear (Fig \ref{fig3}B). However, the longitudinal traction force pattern will remain similar (Fig \ref{fig3}A). This makes sense as the supra-cellular cable does not directly contribute to the longitudinal traction force. In addition, the boundary of the leading front would be significantly rougher without the cable where many single chains of cells emerges (SI, Movie S3). A cable could, at least partially, stabilize a finger that has lost its leader. If we remove the leader cell but keep the cable in an already formed finger, its shape could be partially maintained with the help from the cable (SI, Movie S4). The cable could prevent the finger cells from moving sideward to the open space, though they tend to move in that direction after losing the guidance of the leader (Fig. \ref{fig4}A, \ref{fig4}B). 

In absence of both the leader cell and the cable, no finger can survive for even short periods of time. In fact, if there is an already existing finger, it would be de-stabilized and vanish (SI, Movie S5). Driven by CIL, cells on sides of the fingers move in the direction perpendicular to the finger after removing the cable and the leader cell. This is easy to understand since the sideward direction is the free space for cells on sides of the finger. Without the cable, nothing will prevent the cells from moving sideward. Without the leader cell, the follower cells will not be pulled or attracted to the fingertip and they will not move in the direction parallel to the finger. The polarity in the finger cells will vanish and the cells have a stronger tendency to move sideward (Fig. \ref{fig4}C, \ref{fig4}D.). 

One of the most surprising predictions of our approach is the necessity of modifying the phenotypic properties of not just the leader itself but also the nearby followers. If we abandon the graded behavior from the leader cell to the follower cells by only assigning a larger self-propelled force $m$, maximum cell-cell adhesion, and friction coefficient $\xi$ etc. to the leader cell itself, no finger will emerge (SI, Movie S6). Previous models achieve a similar transferring of the leader cell effects to its followers via a Vicsek interaction \cite{NGovIntBio15, HakimComBio13}. They show that this mechanism may be crucial for the formation of the finger \cite{NGovIntBio15}. Our more detailed model indicates that merely having a large pulling force (enhanced traction and adhesion) between the leader cell and its adjacent neighbors is not enough to guide a sufficient number of follower cells. Instead, we had to include a mechanism to transfer the effects of this pulling force or attraction to cells further behind the leading edge. We have assumed that this information transfer leads to different phenotypic characteristics for the follower cells. This is different than assuming that the difference emerges dynamically among identical cells due to finger geometry modulating mechanical interactions. Our prediction can be tested by carefully analyzing phenotype as a function of spatial position. As mentioned above, initial experimental efforts in this direction have indeed detected a type of graded intermediate phenotype in the follower cells. 

This model predictions we have given here is not sensitive to parameters. The parameters listed in table S1 in SI could vary in a large range without changing the output. We always obtain a finger if we have a leader cell with enough pulling force and follower cells with the graded behavior. It should by now be clear that our interest is not in tuning the parameters to fit some specific experiments. Instead, our method makes direct predictions about necessary and sufficient conditions for the emergence of the patterned structures observed in experiments. 

\section{Front stability via an active matter model} 

As discussed above, the issue of front stability in active model systems and its connections to leader cells has been the subject of recent investigations in the active matter community. Our computational model does not exhibit any instability in the absence of phenotypic variability and in fact shows the necessity of both the leader cell and the graded behavior for the formation of a finger protrusion. To better place this result in context we now consider a continuum version of our model and demonstrate that it to exhibits the same stability characteristics. 

As our hydrodynamical model, we assume a continuous, homogenous and impressible two-dimensional active fluid in our active matter model. The Toner-Tu equations based on these assumptions become
\begin{subequations}
\begin{align} 
\nabla \cdot \vec{u} &= 0 \label{eq1a}, \\
\mu \nabla ^2 \vec{u} - \nabla p + a \vec{u} - b|\vec{u}|^2 \vec{u} &= \rho (\frac{\partial \vec{u}}{\partial t} + \lambda (\vec{u} \cdot \nabla) \vec{u}) \label{eq1b},
\end{align} 
\end{subequations}
where $\bm{u}$ is the velocity field for the active fluid, $\rho$ is the density, $\mu$ is the viscosity, and $p$ is the pressure. The continuous equation \eqref{eq1a} indicates the impressibility, meaning the cells will not proliferate or die. This is a reasonable assumption considering the time scale of the processes we are interested in. In fact, in most wound healing experiments the cell proliferation rate is very low, and the division usually happens in the center of the monolayer, which rarely affects the leading front (This is also true in our computational approach, SI Movie S7). Equation \eqref{eq1b} describes the force balance in the model. The two terms $a \bm{u}$ and $b |\bm{u}|^2 \bm{u}$ account for the activity, where the first term acts as self-propulsion forces with a positive coefficient $a$ and the second term accounts for the friction between the tissue and the substrate with a coefficient $b$. The viscosity and pressure terms are for a Navier-Stokes equation for conventional fluids. The $\lambda$ term in equation \eqref{eq1b} indicates the total momentum is not conserved and our system is not constrained by Galilean invariance. However, considering the low Reynolds numbers of our system, where viscous forces are much larger than inertial forces, we can set the right hand side of equation \eqref{eq1b} to zero. (Without this simplification, we would have an extra term proportional to $(\lambda -1)$  in our equations as well as extra factors of the growth rate; the latter cannot alter the existence or non-existence of a steady-state instability.) We apply this simplification in our following calculations. 
\begin{figure}[htbp] 
\centering
    \includegraphics[width=.65\linewidth]{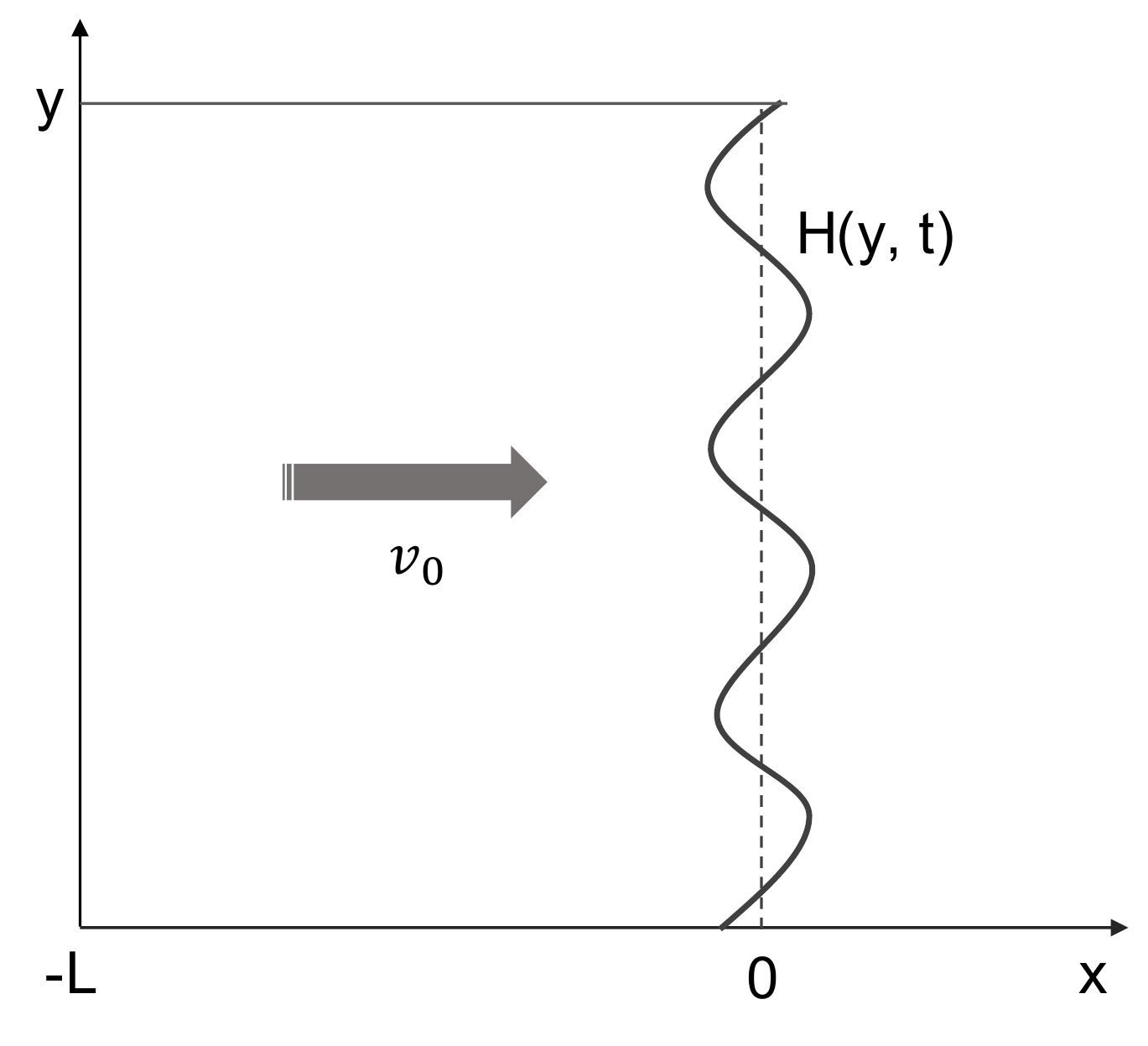} 
  \caption{Sketch for the geometry of the monolayer.}
  \label{fig:InsAF} 
\end{figure}

To describe a monolayer of active fluid, we need to define a geometry for our system with appropriate boundary conditions. We assume that initially we have a strip of active fluid that moves uniformly in the $x$ direction with a speed $v_0$; from equation \eqref{eq1a} and \eqref{eq1b}, we get $v_0 = \sqrt{a/b}$. The cellular region lies between $ x = -L + v_0 t $ and $ x = H(y, t) $ in the x direction, $ y = - \infty $ and $ y = \infty $ in the y direction.  We assume a hard wall at the rear  $ x = -L $ and a free surface on $ x =  H(y, t)$; initially $H=0$. This interface corresponds to the leading front of a collectively migrating monolayer of cells. The hard wall must move at the mean velocity because of the incompressibility condition.In the following section, we analyze the linear stability at the leading front. 

Perturbatively, we have $\vec{u} = v_0 {\hat{x}} + \delta \vec{u}$, $p = p_0 + \delta p$, and $H = v_0 t +  h$. To model the leader cell, we introduce a curvature based force on the leading front, which regulates the parameter $a$. We take the strength of self-propulsion $a$ to be a function of $f$ and  for simplicity, we only consider the first order term, 
\begin{subequations}
\begin{equation}
a(f) = a_0 + f a_1 \label{eqla},
\end{equation}
where $f$ is regulated by
\begin{equation}
\frac{\partial{f}}{\partial{t}} = D_f \nabla ^ 2 f - \alpha f \label{eqlb},
\end{equation}
where $D_f$ and $\alpha$ are constants. On the leading front, we have a boundary condition that $f$ is proportional to the local curvature,
\begin{equation}
f|_{at\ the\ leading\ front}  = q^2  h. \label{eqlc}
\end{equation} 
\end{subequations}

The motivation of this curvature based force is provided by our computational approach presented in the previous section and also from earlier finger-based models \cite{SilberzanBioPMod10}. As already discussed, we show the strong pulling force from the leader cell and a graded behavior in which the neighbors of the leader cell adopt partial leader phenotype are necessary for the formation of stable fingering protrusions. Here, this curvature based force provides similar effects via the diffusion-decay equation \eqref{eqlb}, which, we believe, is the simplest assumption in this active fluid model. In addition, the curvature may be one of the mechanical cues responsible for the emergence of the leader cell. In any case, the leader cells are in fact associated with points of high positive curvature.

The perturbation applied on the leading front can be expressed in normal modes, i.e.
\begin{subequations}
\begin{align} 
u_x &= A e^{r (x - v_0 t) + \omega t + iqy}  \label{eqnma} ,\\
u_y &= B e^{r (x - v_0 t) + \omega t + iqy}  \label{eqnmb} ,\\
p &= C e^{r (x - v_0 t) + \omega t + iqy}  \label{eqnmc} ,\\
h &= h_0 e^{\omega t + iqy} \label{eqnmd},\\
f &= F e^{r (x - v_0 t) + \omega t + iqy} \label{eqnme}, 
\end{align}
\end{subequations}
where $q$ is the wave number and the real part of $\omega$ is the growth rate.  Plugging equations \eqref{eqnma}-\eqref{eqnme} into equations \eqref{eq1a}, \eqref{eq1b} and \eqref{eqlb}, we get a set of linear equations for $A$, $B$, $C$ and $F$ (more details in SI). We can solve for the allowed values of the $r$ from the self-consistency of this set of linear equations. Given the setup and imagining that $L$ is very large, we only keep values of $rf$ with positive real parts and  then ignore the boundary conditions at the rear moving wall. As shown in the SI, there are three such values of $r$. For each of these wave-vectors, we can obtain the relationship between the perturbation coefficients $A$, $B$, $C$, and $F$. The simplest of these is the mode that arises purely from the $f$ equation, which has 
$$
r_3 = \frac{-v_0 + \sqrt{v_0^2 + 4 D_f (D_f q^2 + \alpha + \omega)}}{2 D_f}.
$$
Again, we employ a quasi-static approximation and drop the $\omega$ term instead the square root. Again, this cannot affect whether or not there is a real mode instability (since $\omega$ is obviously irrelevant right where it crosses zero) but could in principle allow us to miss a Hopf bifurcation. 

The following boundary conditions hold at the leading front: 
\begin{eqnarray}
\frac{\partial{h}}{\partial{t}}  &=& u_x \label{eqbca}, \nonumber \\
\frac{\partial{u_x}}{\partial{y}} + \frac{\partial{u_y}}{\partial{x}} & = & 0, \nonumber \\
2 \mu \frac{\partial{u_x}}{\partial{x}} - p & = & \gamma \frac{\partial ^ 2{h}}{\partial{y^2}}. 
\end{eqnarray}
The first refers to the continuity of velocity, the next indicates no transverse stress, and the final equation means that the normal stress equals the surface tension. With these boundary conditions, we can solve for the growth rate $\omega$.
Plugging $v_0 = \sqrt{a_0/b}$ and equations (S2) in SI into the boundary conditions, yields a set of linear equations, from which $\omega$ can be determine as a function of $q$. 
\begin{figure}[htbp] 
\centering
    \includegraphics[width=.95\linewidth]{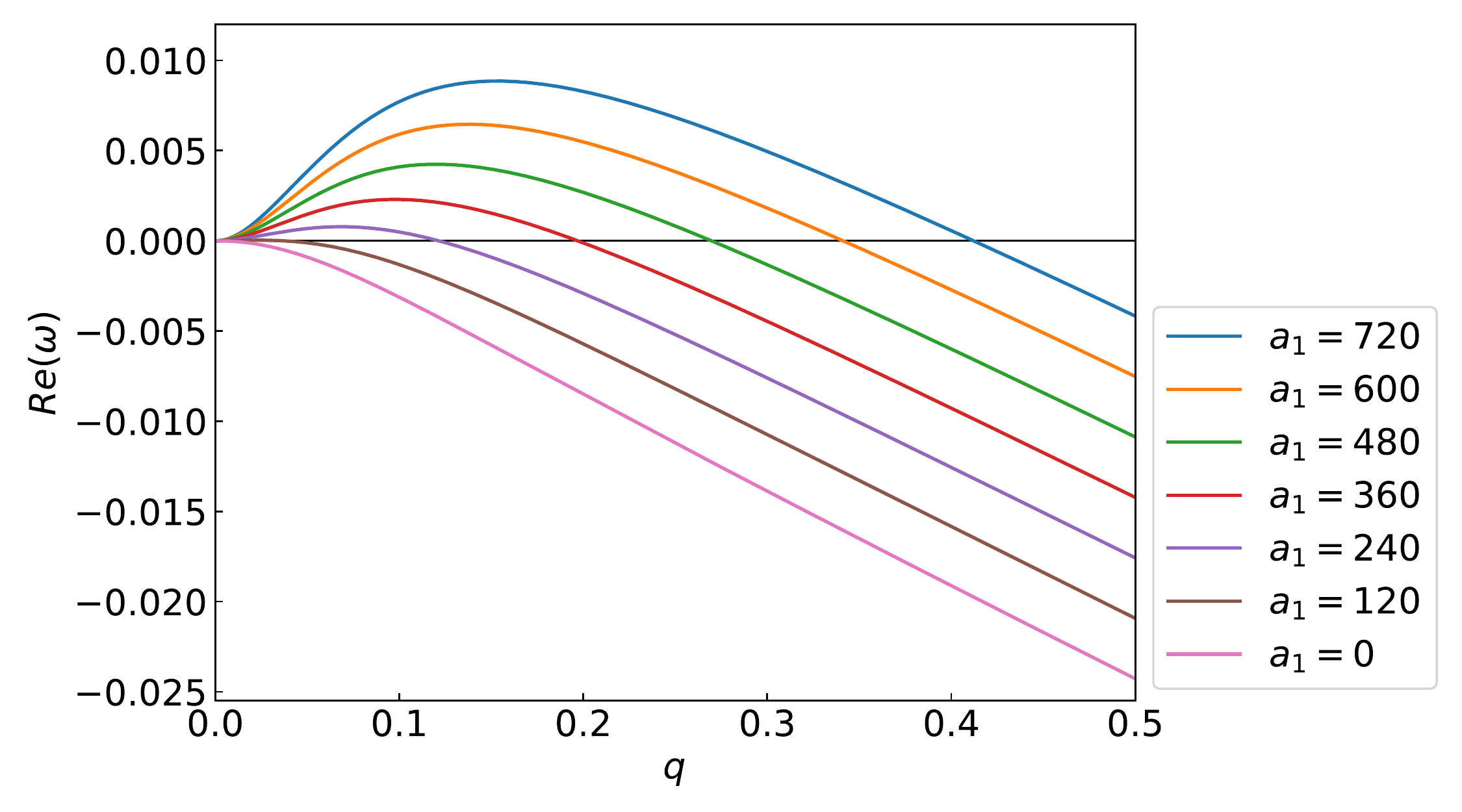} 
  \caption{Real part of the growth rate $\omega$ as a function of wave number $q$ for the moving case. $a_1$ is a parameter in equation \eqref{eqla}, where a larger $a_1$ corresponds to a larger curvature based force on the leading front. }
  \label{fig:InsOmegaQ} 
\end{figure}
Assuming values for the other constants (more details in SI), we can obtain plots of $Re(\omega)$ and $q$ (Fig. \ref{fig:InsOmegaQ}). Only when $a_1 > 0$, will we have a positive $\omega$. Therefore, only when there is a extra force at the leading front, can the instability emerge. When $a_1 = 0$, $\omega$ is always negative, which agrees with \cite{LeePRE17}. This finding is consistent with the result obtained in our computational model that the absence of leader cells leads to a flat front. It is worth noting however that one can obtain instabilities by adding in additional physical effects such as polarization, growth, or interaction of the front with a different non-trivial cellular layer. This will be discussed below.

\section{Discussion}
We have developed and performed simulations of a computational model to understand the formation of fingers during the spreading of cellular monolayers on a stiff substrate. Our computational approach combines mechanical aspects of collective cell motility, as has been developed previously based on a subcellular-element method, as well as the added possibility of phenotypic variability. This variability takes the form of leader cells and their effects on their immediate neighbors and also the form of boundary cells that connect together to form a cell-spanning cable. We explicitly show that within our model the leader cells and graded follower behavior are necessary to form a finger. In addition to the computational approach, we have developed an active fluid model for the stability analysis on the leading interface. In the absence of growth effects or of strong collective polarization (as would be the case if CIL is more dominant in controlling cell polarity), at least some form of long-range force on the leading front is required for the emergence of a fingering protrusion. Our active fluid model provides a way to incorporate a curvature-based force in a Toner-Tu style hydrodynamics, which approximately corresponds to the extra pulling force from the leader cell as well as the graded follower behavior as used in our computational approach. This stability study therefore supports the assumptions made in our computational model. 

Past work has addressed the possibility of forming fingers without a leader cell \cite{BasanPNAS13, SavinRSOS18}. Basan et al. provided a subcellular-element-based model with a strong velocity alignment among nearby cells \cite{BasanPNAS13}. This alignment accounts for the fingering protrusions seen in their model. However, if we introduce contact inhibition of locomotion (CIL) in this model, the fingering protrusions vanish if there is no leader cell \cite{ZimmermannPNAS16}. This change in the model was occasioned by several studies which argued that CIL was necessary in order to provide robust explanations within our modeling framework for data arising from a variety of experimental protocols. Bogdan et al. also suggest a fingering instability without any leader in their active fluid model \cite{SavinRSOS18}. The major difference between their continuum model and ours is that they have a positive net rate of growth while ours is zero (right hand side of equation \eqref{eq1a}). They show that fingering protrusions could emerge in a circular geometry with this large proliferation rate. In our computational model, the interface spreads mostly by active motility as opposed to being pushed by growth; the latter would give rise to a pressurized interior in contrast to what is observed.   Alert et al. show that the fingering instability could be formed in a spreading epithelia based on a kinematic mechanism using an active polar fluid model \cite{AlertPRL19}. Their treatment of growth is to assume that the active medium is highly compressible, meaning in detail that they do not have the impressibility equation or a pressure term. In the SI, we show that this compressibility is not a critical factor in the difference between their result and ours. We can therefore conclude that the polarity field in their model introduces a velocity alignment, which results in a velocity gradient from the leading front to the back. This velocity gradient accounts for the interface instability. Actually, this polar field acts similarly to velocity alignment with neighboring cells as in the particle based model developed by Basan et al.\cite{BasanPNAS13}. This instability can therefore be expected to be stabilized by strong enough CIL. Given that CIL seems to be a necessary part of any model which correctly reproduces the observed mechanical stress in the bulk of steadily expanding tissue, we expect that the monolayer of epithelia tends to have a stabilized  interface without any additional asymmetric force. It is clear, though, that the full role of CIL in interfering with stable fingering deserves additional study. 

As emphasized throughout, our models addressed the necessity and the role of the leader cell. The leader cell, once it emerges, will have a large self-propelled force. Meanwhile, it moves in the direction determined by CIL, which makes it appear that the leader cell is ``guided" by its follower cells; this is perhaps similar to ideas presented by ref \cite{ SpatzNC18} which argues that the initial stages of finger formation  are mostly driven by follower cell dynamics. It is equally valid to state that the fast moving leader cell invades the free space and the follower cells move after it and indeed collective motility certainly means that the behavior is determined collectively.  In our model, the leader cell guides the follower cells through two ways: (1) a strong pulling force directly acts on the follower cells which are its immediate neighbors; (2) the graded phenotypic variation from the leader cell to follower cells behind it affect the parameters associated to those cells. The first effect is straightforward to appreciate and could be achieved through cell-cell junctions. However, within our model this effect on its own is not enough to form a finger. Therefore, we proposed the additional mechanism in which the leader cell alters nearby cells within a certain range; this is partially based on experimental observations of an intermediate phenotype \cite{SilberzanPNAS07, SilberzanNCB14}. It is also consistent with other observations including an increased number of focal adhesions on the leading edge, a large number of E-Cadherins between cells in the finger etc \cite{SilberzanPNAS07}. Within this range, cells are taken to have an additional self-propulsion force in the direction towards the leader cell and a larger adhesion force connecting them to their neighbors. This graded behavior facilitates the formation of the finger and leads directly to the polarity of cells within the finger \cite{SilberzanBPOP11}. In fact, experimental observations show that this polarity will vanish if the leader cell is removed \cite{SilberzanBPOP11}. This long range guidance might be a result of biochemical cues from the leader cell as passed on via the strengthened links between the follower cells, i.e. the phenotype responds to force and eventually establishes the graded behavior self-consistently. Alternatively, there might be specific biochemical signals originating at the leader cell that decay as we move further back into the finger. More generally, there is in all probability a complex feedback between different mechanical and biochemical signals. One clue as to what could be occurring emerges from the work on the role of Notch-Delta signaling in the fingering process \cite{WongNC15}. Also,  as already mentioned, there is direct evidence \cite{MarcusNC17} regarding leader cells in  a 3d context that they secrete diffusible chemical signals such as VEGF which modulate follower cell motility.We leave further exploration of this important issue for future research.

Our simulation predicts that a finger can be formed without any supracellular cable. Instead, the major role of the observed cable is to maintain the shape of the finger. Without the cable, the boundaries on both sides of the finger become significantly rougher. We also predict that the transverse traction force pattern would be markedly different without the cable and would not match  the one seen in experiments where in fact a  supracellular cable is observed on sides of a finger \cite{SilberzanPNAS07, SilberzanNCB14}. The supracellular actomyosin cable is also observed in other scenarios, for example, on the boundary of a localized wound. Previous work has described the role of this cable during the healing of a circular wound \cite{YangSM18}, as it is necessary to have the supracellular cable on the edge to close the circular wound when the cells cannot crawl on the substrate \cite{YangSM18, LadouxNC15, LadouxNM14}. The cable also keeps the wound boundary smooth. 

This work should motivate the modeling community to take into account cell-to-cell differences when constructing approaches to tissue morphology and motility. While it is always simpler to imagine a uniform set of active particles giving rise to an effective hydrodynamics with uniform spatial parameters, the utility of this simplification for biological systems is, in our opinion, limited. Cells are not colloids, but instead actively sense their surroundings and adjust their interactions accordingly. Here we have argued that this is crucial for finger formation and we expect that it will be crucial for many other examples of collective cell behavior. Our phenomenological approach has taken this phenotypic variability as given and  determined the mechanical consequences thereof. More complete models to be developed in the future will determine cell phenotype self-consistently in concert with determining mechanical behavior, taking into account the reciprocal interactions between these aspects of real biological cells.

\acknowledgments{This work was supported by the National Science Foundation Center for Theoretical Biological Physics (NSF PHY-1427654). We thank David Kessler for useful discussions. }

\bibliographystyle{hunsrt}

\end{document}


\beginsupplement
\maketitle
\section*{Details for the computational model}
\subsection*{Dynamics in our model}
In our model, each cell is represented by two subcellular elements (particles), the front and the rear particle. Each cell has a polarity which is defined by $\bm{r} = \bm{r}_f - \bm{r}_r$, where $\bm{r}_f$ and $\bm{r}_r$ are the positions for the front and rear particles respectively. The self-propulsion force is assigned to each particle, $\bm{m}_f$ for the front particle and $\bm{m}_r$ for the rear particle. The direction for the propulsion force is along $\bm{r}$ but pointing to opposite directions for the front and rear particles for an isolated cell. For cells in a cluster, both the direction and magnitude of the self-propulsion force is regulated by contact inhibition of locomotion CIL (see later). The self-propulsion force is balanced by the intracellular contraction force $\bm{f}_{contr} = (-f_{contr}^0 r/(R_{contr} - r) + f_{exp}/r)\bm{\hat{r}}$ (use $f_{contr}^l$ for leader cells and $f_{contr}^b$ for boundary cells). This force is attractive for most distance r and only repulsive at extremely short distances, the latter of which simulates a hard core. 

The intercellular force between particles from different cells is modeled by  $\bm{f}_{adh/rep}(\bm{r}) = (-A(B-r)+C(B-r)^3)\hat{\bm r}$ for distances within $R_{cc}$ and zero further away. For distance between $R_{rep}$ and $R_{cc}$, the force is attractive which simulates the cell-cell adhesion. For distance below $R_{rep}$, the force is repulsive which models the volume exclusion of cells. To account for cells of different lengths, we adjust the units using the length of the longer cell $l$. We assume $R_{cc} = 1.3l$ and $R_{rep} = 0.75 R_{cc} = 0.975l$. We also assume a fixed maximum adhesion value $f_{adh}^{max}$ (the minimum of $f_{adh/rep}(r)$, we use different values for interaction between different kind of cells). Then we have, $B = R_{cc}$, $A = f_{adh}^{max} 3^{3/2}/(2(R_{cc}-R_{rep}))$, and $C=A/(R_{cc}-R_{rep})^2$. For cells shorter than a minimum value $l_{min} = R_{cc}^{min}/1.3$, we assume $R_{cc} = R_{cc}^{min}$. For cells longer than a maximum value $l_{max} = R_{cc}^{max}/1.3$, we assume $R_{cc} = R_{cc}^{max}$.

The friction between cells and the substrate is $\bm{f}_{fric} = \xi \bm{v}$, where $\xi$ is a constant for each particle. We assume a stiff substrate and the friction balances the propulsion force and traction force. Therefore, we can determine the traction force $\xi \bm{v} - \bm{m}$ in our model.

Motile and non-motile cells are distinguished in our model. For a motile cell, we have $m_f > m_r$, while for a non-motile cell, we set $m_f = m_r$. Both $m_f$ and $m_r$ are constants in our model. Cells can switch between motile and non-motile states. A non-motile cell can transition to a motile cell at a fixed probability $k_+$, and the choice of front and rear particle is random with an equal chance. A motile cell can also become a non-motile cell at a probability depending on the alignment between cell polarity ($\bm{r}$) and its time-averaged velocity ($\bm{v_m}$), $k_- = k_-^0 exp(-c_{trans} \bm{\hat{v}}_m \cdot \bm{\hat{r}})$, where both $\bm{r}$ and $\bm{v}_m$ were normalized, i.e. $\bm{\hat{r}} = \bm{r}/|r|$ and $\bm{\hat{v}}_m = \bm{v}_m/|v_m|$.

The contact inhibition of locomotion (CIL) regulates the self-propulsion force. To calculate the self-propulsion force for each particle i, we first calculate the sum of normalized vector connecting this particle i and its neighboring particles j's (including its partner particle in the same cell) within a distance $R_{inh}$, i.e. $\bm{R}_i = \sum_{j, r_{ij}<R_{inh}} \bm{\hat{r}}_{ij}$. The self-propulsion force becomes $\bm{m} = -m \bm{R} /n$, where $m$ is $m_f$ or $m_r$, n is the number of its neighbor particles. We assume $R_{inh} = R_{cc}$ in our model. 

\subsection*{The leader cell and its effects}
The leader cell has a different morphology and properties. It is selected randomly from cells on the leading front. Once selected as a leader cell, the cell will have all properties of the leader cell and it will prevent its neighboring cells within a certain range from being selected as a new leader cell. Both particles of the leader cell will gain an additional self-propulsion force $\Delta m$ and its direction will be determined by CIL. The coefficient for the intracellular interaction will be increase to $f_{contr}^l$ and the maximum adhesion/repulsion value will be increased to $f_{adh}^{max}$. There will be an additional friction coefficient $\Delta \xi$ and a larger threshold of division which simulate the larger size of the leader cell. 

The leader cell also affects its neighbors within a certain distance $R_{LC}$, which is called the graded behavior. All cells within this range will have an additional self-propulsion force $m_{extra}$ pointing to the center of the leader cell. The magnitude of this additional self-propulsion force is given by $m_{extra} =  m_{max} \cdot (exp(-D \cdot r/R_{LC}) - exp(-D)) / (1.0 - exp(-D))$, where $m_{max}$ is the maximum additional propulsion force, $D$ is the decay rate and $r$ is the distance between the particle and the leader cell. The coefficient of the intercellular interaction, the maximum adhesion/repulsion value, the division threshold and the friction coefficient will increase in a certain amount defined by the similar equation, where we can change $m_{extra}$/$m_{max}$ to the corresponding increase parameters. We check the eligibility of the leader cell at every step. If there are less than 6 particles behind the leader cell (the leader cell detaches from the cellular sheet) or more than 6 particles in front of it (the leader cell is no longer leading), the leader cell will turn back to a regular cell and a new leader cell might emerge. In this way, we develop a model for the intermediate phenotype between the leader cell and the follower cells far away.

\subsection*{The supracellular actomyosin cable}
\begin{figure}[htbp] 
\centering
    \includegraphics[width=.8\linewidth]{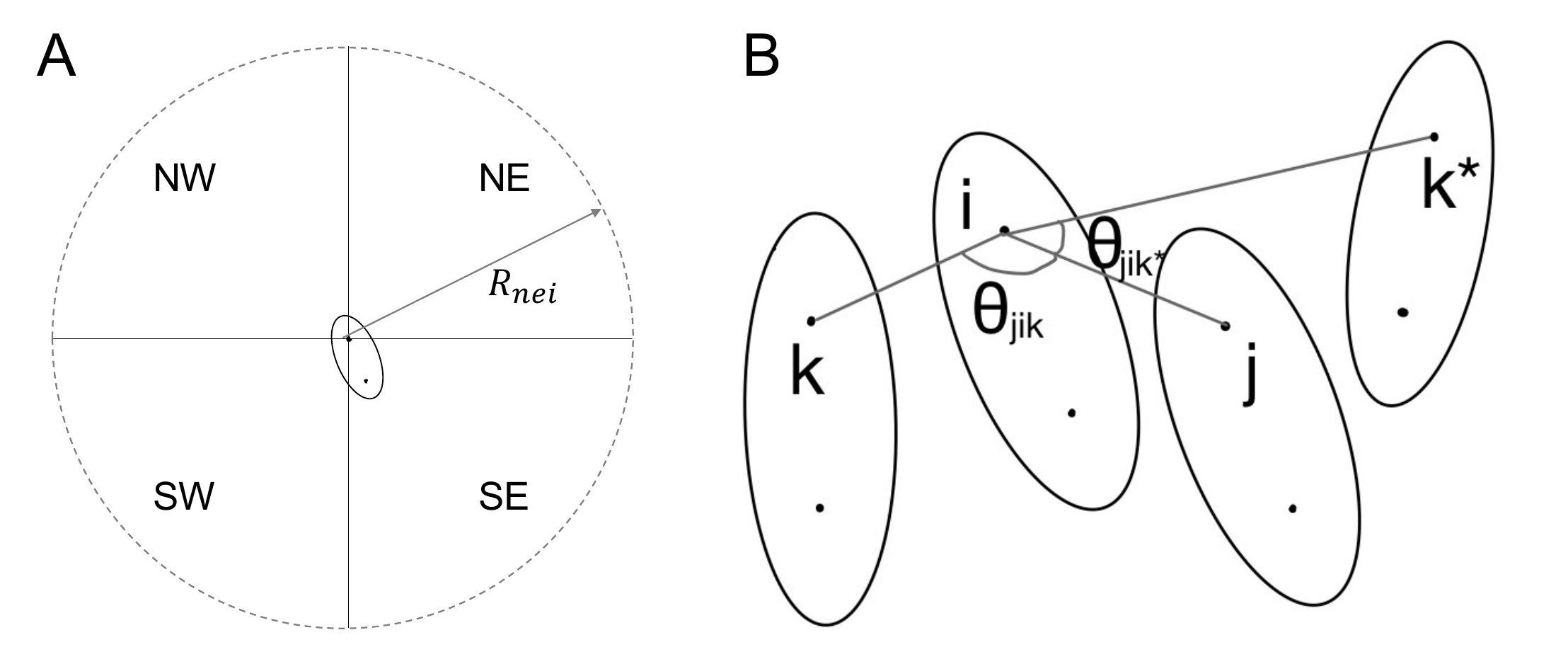}     
  \caption{Boundary cells. A. Subregions around a sub-cellular element. B. Example for deciding unhidden nearest neighbors, where k is an unhidden nearest neighbor of i and $k^*$ is not.}
  \label{figS1} 
\end{figure}

Cells on the leading front except the fingertip will be connected through the supracellular actomyosin cable. This cable connects the cells on the boundary mechanically. We develop an algorithm to detect the boundary cells. For each particle, we count the number of other particles within a range of $R_{nei}$. In Cartesian coordinates, we divide this range into four subregions northeast (NE), southeast (SE), southwest (SW), northwest(NW) and count the total of particles in each subregion (Figure S1A). If the total particles in one subregion is less than the average number of total particles in other three subregions by 5 or the total particles in two adjacent subregions is less than the total particles in the other two adjacent subregions by 8, we designate the cell containing this particle as a `boundary cell'. If a particle from a `boundary cell' picks up a balanced number of surrounding cells in all four subregions, i.e. the difference between the total particle number in any subregion and the average number of total particles in the other three subregions is less than 2, the cell will transit back to a regular cell.

To construct the supracellular actomyosin cable, we connect particles from boundary cells with each other's unhidden nearest neighbors. To find out each particle's unhidden nearest neighbors, we first rank particles (excluding the partner particle) around particle i based on their distance to particle i. Then we find the nearest particles that is not hidden by other nearest particles by applying the following procedure: (1) Calculate the angel $\theta_{jik}$ between the line connecting the new particle k and i and the line connecting an already selected particle j and i. (2) If for all j, $\theta_{jik}$ is larger than $90$ degrees, we select particle k as particle i's unhidden nearest neighbor. (3) Repeat this procedure until we find all the unhidden nearest neighbors (Figure S1B). After finding all the nearest neighbors for each boundary cell, we apply an additional intercellular force between these particles, $\bm{f}_{cable} = (-f_{cabcontr}^0 r/((R_{cabcontr} - r) r) + f_{cabexp}/r)\bm{\hat{r}}$. This additional intercellular force behaves like the intracellular contractive force; it is attractive for most distances and repulsive only for extremely small distances. 

\section*{Supplementary information for the active matter model }
\subsection*{Stability analysis}
Plug equations (4) in the main article into equations (2) and (3b), we get a set of linear equations for A, B, C and F, 
\begin{subequations}
\begin{align}
r A + i q B = 0 \label{eqpa}, \\
\mu (r^2 - q^2) A - r C + a_0 A + a_1 v_0 F - 3 b v_0^2 A = 0 \label{eqpb}, \\
\mu (r^2 - q^2) B - i q C + a_0 B - b v_0^2 B = 0 \label{eqpc}, \\
(\omega - v_0 r) F = D_f (r^2 - q^2) F - \alpha F \label{eqpd}.
\end{align}
\end{subequations}
Assuming $F \rightarrow 0$, from equations \eqref{eqpa} - \eqref{eqpc}, we can get four solutions of $r$, 
$$
r_{1/2} =  \pm \sqrt{q^2 \pm \sqrt{-2 q^2 a_0/\mu}}.
$$
Assuming $F$ is non-zero, we can get another solution for $r$ from equation \eqref{eqpd}, 
$$
r_3 = \frac{-v_0 \pm \sqrt{v_0^2 + 4 D_f (D_f q^2 + \alpha + \omega)}}{2 D_f}, 
$$
which corresponds to the modes regarding the curvature based force we introduce above. Since we consider a rightward moving fluid, we only keep the positive $r's$. Therefore, there will be three modes, and we can obtain
\begin{subequations}
\begin{align}
u_x &= \sum_{i=1}^2 A_i e^{r_i (x - v_0 t) + \omega t + iqy} + F_A h_0 e^{r_3 (x - v_0 t) + \omega t + iqy} \label{eqnmsa},\\
u_y &= \sum_{i=1}^2 B_i e^{r_i (x - v_0 t) + \omega t + iqy} + F_B h_0 e^{r_3 (x - v_0 t) + \omega t + iqy} \label{eqnmsb},\\
p &= \sum_{i=1}^2 C_i e^{r_i (x - v_0 t) + \omega t + iqy} + F_C h_0 e^{r_3 (x - v_0 t) + \omega t + iqy} \label{eqnmsc},\\
h &= h_0 e^{\omega t + i q y} \label{eqnmsd}.
\end{align}
\end{subequations}

From equations \eqref{eqpa} - \eqref{eqpd}, we can express $B$ ($F_B$), $C$ ($F_C$) in $A$ ($F_A$)  and the expression of $F_A$, i. e. 
\begin{subequations}
\begin{align}
B &= -\frac{r}{i q} A ,\\
C &= \frac{r}{q^2} (\mu (r^2 - q^2) + a_0 - b v_0^2) A , \\
F_A &= -\frac{a_1 v_0 q^2}{(1 - r_3^2/q^2)(\mu (r_3^2 - q^2) + a_0 - b v_0^2) - 2 b v_0^2} , \\
F_B &= -\frac{r}{i q} F_A ,\\
F_C &= \frac{r}{q^2} (\mu (r^2 - q^2) + a_0 - b v_0^2) F_A .
\end{align}
\end{subequations}
Using the above relationship and equation \eqref{eqnmsa} - \eqref{eqnmsd} and the boundary conditions, we can get a set of linear equations for the coefficients. 

We plug in $v_0 = \sqrt{a_0/b}$ and the normal modes expressions (substitute the coefficients in $A$ or $F_A$) into the boundary conditions, we get a set of linear equations.   
\begin{subequations}
 \begin{align}
 A_1 + A_2 - (\omega - F_A) h &= 0 ,\\
 \frac{q^2 + r_1^2}{q} A_1 + \frac{q^2 + r_2^2}{q} A_2 + \frac{q^2 + r_3^2}{q} F_A h &= 0 ,\\
 \frac{\mu r_1 (3 q^2 - r_1^2)}{q^2} A_1 + \frac{\mu r_2 (3 q^2 - r_2^2)}{q^2} A_2 + \nonumber \\ (\gamma q^2 + \frac{\mu r_3 (3 q^2 - r_3^2)}{q^2} F_A) h &= 0 .
 \end{align}
\end{subequations}
Since $A_1$, $A_2$ and $h$ are non-zero, we have a characteristic equation, 
\begin{equation}
det \begin{bmatrix} 1 & 1 & - (\omega - F_A) \\ \frac{q ^ 2 + r_1 ^ 2}{q} & \frac{q ^ 2 + r_2 ^ 2}{q} & \frac{(q ^ 2 + r_3 ^ 2) F_A}{q} \\ \frac{\mu r_1 (3q ^ 2 - r_1 ^ 2) }{q ^ 2} & \frac{\mu r_2 (3q ^ 2 - r_2 ^ 2) }{q ^ 2} & \gamma q ^ 2 + \frac{\mu r_3 (3q ^ 2 - r_3 ^ 2) F_A }{q ^ 2}  \end{bmatrix} \quad =\quad 0. 
\end{equation} 
For simplicity, we drop the $\omega$ term in $r_3$ when solving this characteristic equation. This approximation is reasonable since the $\omega$ is near $0$ in scenarios we are interested in and the time duration is sufficiently short. Therefore, we get the expression of $\omega$,
\begin{equation}
\omega = F_A 
- \frac{\frac{q^2}{\mu}(r_2^2 - r_1^2)(\gamma q^2 + \frac{\mu r_3}{q^2}(3 q^2 - r_3^2) F_A) + (q^2 + r_3^2)F_A(3 q^2 (r_2 - r_1) - (r_2^3 - r_1^3))}{(r_1 - r_2)(3 q^4 - q^2 r_1^2 - 4 q^2 r_1 r_2 - q^2 r_2^2 - r_1^2 r_2^2)}, \label{eqomega}
\end{equation}
where $F_A$ and $r's$ can be expressed in $q$. Assuming values for the constants, we can obtain plots of $Re(\omega)$ and $q$ (Fig. 7). The constants used in Fig. 7 are $\mu = 10 ^ 4~ \si{\Pa~\s}$, $\gamma = 10 ^ 3~ \si{\Pa~\mu m}$, $L = 100~ \si{\mu m}$, $a_0 = 60~ \si{\Pa~\s~\mu m ^ {-2}}$, $b = 100~ \si{\Pa~\s^3~\mu m^{-4}}$, $D_f = 10 ^ 4~ \si{\mu m^{2}~ s^{-1}}$, $\alpha = 10~\si{s^{-1}}$.

\subsection*{Comparison with active polar fluid model}
\subsubsection*{Remove incompressibility in our model}
To compare with Alert et al.'s active polar fluid model where they did not have the incompressibility and pressure, we modify our equation (1),
\begin{equation}
\mu \nabla ^2 \vec{u}  + a \vec{u} - b|\vec{u}|^2 \vec{u} = \rho (\frac{\partial \vec{u}}{\partial t} + \lambda (\vec{u} \cdot \nabla) \vec{u}) \label{eqsneq1}.
\end{equation}
The right hand side is set to $0$ as assumed in the main article. Assuming a similar normal modes for $u_x$ and $u_y$, we get, 
\begin{subequations}
\begin{align}
\mu (2 r^2 - q^2) A  + \mu i q r B  + a_0 A  - 3 b v_0^2 A = 0 \label{eqsnpb}, \\
\mu (r^2 - 2 q^2) B + \mu i q r A + a_0 B - b v_0^2 B = 0 \label{eqsnpc} .
\end{align}
\end{subequations}
Then we can solve for the relationship between $A$ and $B$, 
\begin{equation}
B = \frac{i}{\mu q r}[(\mu (2 r^2 - q^2) - 2 a_0] A .
\end{equation}
Then $r$ can also be determined, 
\begin{equation}
r_{1/2} = \sqrt{\frac{(2 q^2 + a_0) \pm \sqrt{a_0^2 - 4 a_0 q^2}}{2}} .
\end{equation} 
There are two allowed modes. Similarly, $u_x$, $u_y$ and $h$ can be expressed in normal modes. Plug them in the boundary conditions in equation (6) in the main article (removing the $p$ term), yields a set of linear equations, 
\begin{subequations}
\begin{align}
A_1 + A_2  - \omega h &= 0 ,\\
(\mu r_1^2 - a_0) A_1 + (\mu r_2^2 - a_0) A_2 &= 0 ,\\
2 \mu r_1 A_1 + 2 \mu r_2 A2 + \gamma q^2 h &= 0 .
\end{align}
\end{subequations}
Solving its characteristic equation yields the growth rate,
\begin{equation}
\omega = - \frac{(r_1 + r_2) \gamma q^2}{2 (\mu r_1 r_2 + a_0)} ,
\end{equation}
which is always negative, meaning it is always stable at the leading front for the moving case. Therefore, removing the incompressibility and pressure term will not lead to an edge instability. 

\subsubsection*{Adding surface tension in Alert et al.'s model}
On the other hand, to check their model, we add a surface tension and remove the $T_0$, $p$ terms in Alert et al.'s model. (They checked the surface tension in their work, we just do it again in our way for completeness.) Their equation (S45) becomes, 
\begin{subequations}
\begin{align}
\eta (2 k^2 - q^2 - \frac{1}{\lambda ^2}) \delta \tilde{v}_x + i q k \eta \delta \tilde{v}_y &= 0 ,\\
i q k \eta \delta \tilde{v}_x + \eta (k^2 - 2 q^2 - \frac{1}{\lambda ^2}) \delta \tilde{v}_y &= 0 .
\end{align}
\end{subequations} 
Solving the characteristic equation, we get,
\begin{subequations}
\begin{align}
k_1 &= \sqrt{q^2 + \frac{1}{\lambda ^ 2}} ,\\
k_2 &= \sqrt{q^2 + \frac{1}{2 \lambda ^ 2}} .
\end{align}
\end{subequations}
From their equation (S44), we can get another solution for k,
\begin{equation}
k_3 = q .
\end{equation}
Assuming $L \rightarrow \infty$ and a surface tension at $x = 0$, then their equation (S47) becomes, 
\begin{subequations}
\begin{align}
&\delta \tilde{v}_x(-\infty) = 0,   &\delta \tilde{\sigma}_{xx} (0) = - \gamma q^2 \delta \tilde{L}, \\
&\delta \tilde{v}_y(-\infty) = 0,   &\delta \tilde{\sigma}_{xy} (0) = 0 .
\end{align}
\end{subequations}
$\delta \tilde{v}_x$ and $\delta \tilde{v}_y$ can be expressed in normal modes, 
\begin{subequations}
\begin{align}
\delta \tilde{v}_x = \sum_{j = 1} ^ 3 A_j e^{k_i x + i q y} , \\
\delta \tilde{v}_y = \sum_{j = 1} ^ 3 B_j e^{k_i x + i q y} .
\end{align}
\end{subequations}
Plug them into the boundary conditions yields,
\begin{subequations}
\begin{align}
2 \eta k_1 A_1 + 2 \eta k_2 A_2 + 2 \eta k_3 A_3 &= -\gamma q^2 \delta L ,\\
k_1 B_1 + k_2 B_2 + k_3 B_3 + i q A_1 + i q A_2 + i q A_3 &= 0 .
\end{align}
\end{subequations}

\begin{figure}[htbp] 
\centering
    \includegraphics[width=.55\linewidth]{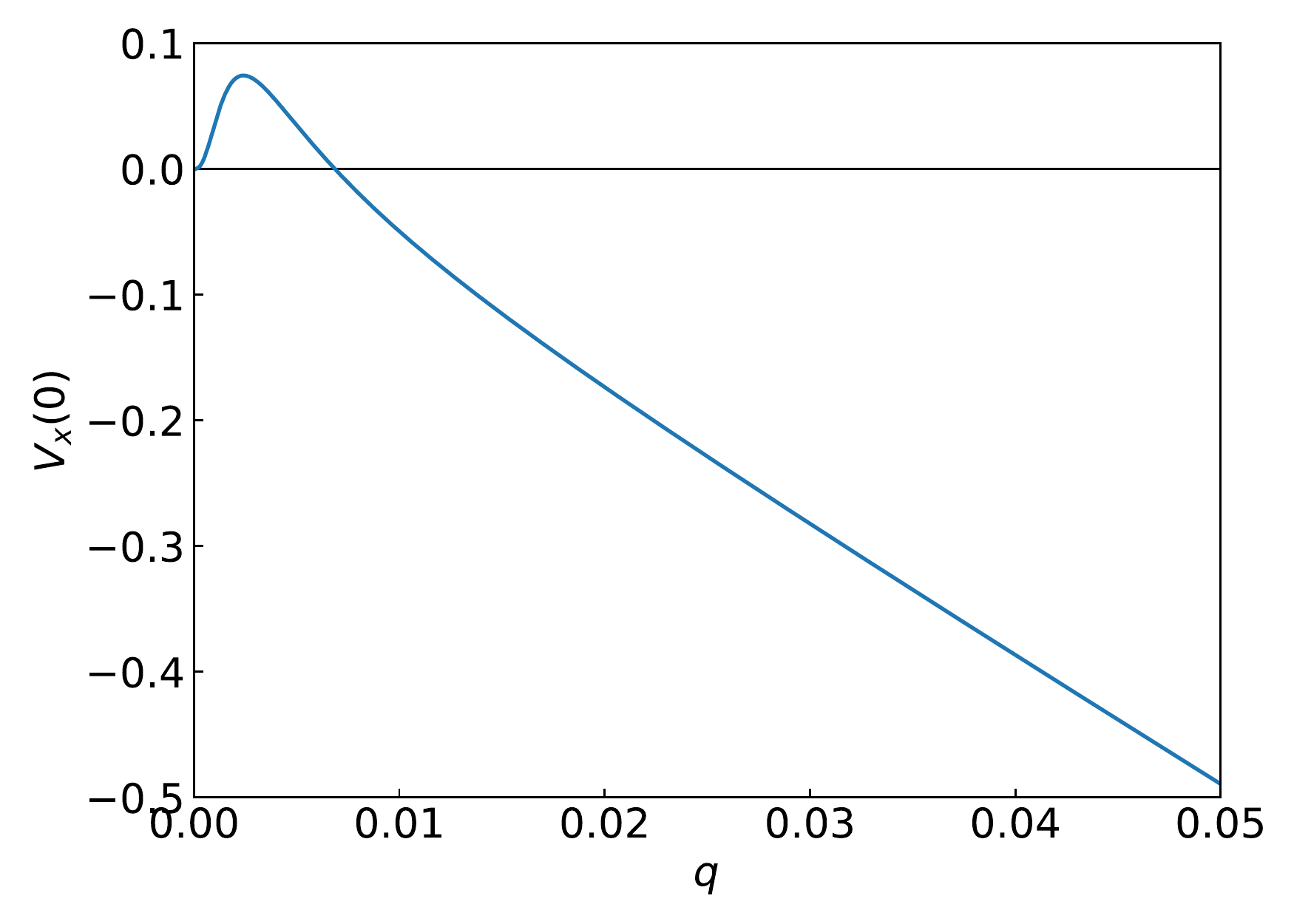} 
  \caption{The x component of the velocity at the leading front $V_x(0)$ as a function of wave number $q$. }
  \label{fig:InsSurTenVx} 
\end{figure}
The relationship between $A$ and $B$ is indicated from equation (S15), 
\begin{equation}
B = \frac{i}{q k} (2 k^2 - q^2 - \frac{1}{\lambda ^2}) A \label{eqsbina}.
\end{equation}
Meanwhile, we can solve for $A_3$ from their equation (S45) and (S46),
\begin{equation}
A_3 = \frac{T_0 q^3 \lambda ^4}{h \eta} \delta \tilde{L} .
\end{equation}
Put everything together, we get $A_1$ and $A_2$,
\begin{subequations}
\begin{align}
A_1 &= \frac{-\frac{\gamma q^4}{\eta} \delta \tilde{L} + \frac{\lambda ^4 T_0 q^3}{h \eta} \delta \tilde{L} (-2 q^3 + 2 q^2 k_2 - \frac{k_2}{\lambda ^2})}{2 q^2 k_1 - 2 q^2 k_2 - \frac{k_2}{\lambda ^2}} ,\\
A_2 &= -\frac{\lambda ^4 T_0 q}{2 h \eta} \delta \tilde{L} (2 q^2 - \frac{1}{\lambda ^2}) - (1 + \frac{1}{2 q^2 \lambda ^2}) A_1 .
\end{align}
\end{subequations}
Therefore, we get the velocity at the leading front, 
\begin{equation}
\delta \tilde{v}_x(0) = A_1 + A_2 + A_3 , 
\end{equation}
which is bounded from above (Fig. \ref{fig:InsSurTenVx}). In this case, the growth rate $\omega$ is also bounded from above, but it still could be positive at small $q$. Therefore, by adding a surface tension will not stabilize their model, which also agrees with their conclusion.

\newpage
\section*{Parameters for the computational simulation}
\begin{table}[htbp]
\footnotesize
\caption{Simulation parameters in simulation units used in Figure 2 and 3.}
\label{tab1}
\begin{tabular}{l c c l}

 \hline
 Parameter & Value & Unit  & Meaning \\
 \hline
 $m_f$ & 1.5 & $p_0 l_0$ &  Magnitude of propulsion force of front particle for motile cells\\
 $m_f$ & 1.3 & $p_0 l_0$ &  Magnitude of propulsion force of front particle for nonmotile cells\\
 $m_r$ & 1.3 & $p_0 l_0$ &  Magnitude of propulsion force of rear particle\\
 $m_{max}$ & 3.5 & $p_0 l_0$ &  Magnitude of additional propulsion force  of the leader and graded behavior\\
 $D$ & 4.0 & & Decay rate for the graded behavior\\
 $f_{exp}$ & 0.02 & $p_0 l_0^3$ & Cellular expansion coefficient\\
 $f_{contr}^0$ & 0.9 & $p_0 l_0^2$ & Cellular contraction coefficient for regular cells\\
 $f_{contr}^b$ & 2.2 & $p_0 l_0^2$ & Cellular contraction coefficient for boundary cells\\
 $f_{contr}^l$ & 3.1 & $p_0 l_0^2$ & Cellular contraction coefficient for leader cells\\
  $f_{contr}^{ext}$ & 2.2 & $p_0 l_0^2$ & Cellular contraction coefficient for the graded behavior\\
 $R_{contr}$ & 2.5 & $l_0$ & Maximum distance for intracellular contraction\\
 $k_+$ & 0.1 & $1/t_0$ & Transition rate to motile state\\
 $k_-^0$ & 0.1 & $1/t_0$ & Transition rate to nonmotile state for zero velocity\\
 $c_{trans}$ & 5.0 & & Scaling parameter for transition to nonmotile state\\
 $f_{adh}^{max(00)}$ & 0.7 & & Maximum cell-cell adhesion between two regular cells\\
 $f_{adh}^{max(0b)}$ & 4.2 & & Maximum cell-cell adhesion between a regular cell and a boundary cell\\
 $f_{adh}^{max(bb)}$ & 4.8 & & Maximum cell-cell adhesion between two boundary cells\\
 $f_{adh}^{max(lx)}$ & 5.6 & & Maximum cell-cell adhesion between a leader cell and another cell\\
  $f_{adh}^{max(ext)}$ & 4.2 & & Additional maximum cell-cell adhesion for the graded behavior\\
 $f_{cabcontr}^0$ & 5.0 & $p_0 l_0$ & Cable contraction coefficient for the supracellular cable\\
 $f_{cabexp}$ & 0.12 & $p_0 l_0^3$ & Cable expansion coefficient\\
 $R_{cabcontr}$ & 3.5 & $l_0$ & Maximum distance for supracellular cable contraction\\
 $R_{cc}^{min}$ & 0.7 & $l_0$ & Minimum interaction range for small cells\\
 $R_{cc}^{max}$ & 1.1 & $l_0$ & Maximum interaction range for large cells\\
 $R_{cc}$ & 1.3$l$ & $l_0$ & Range of adhesive intercellular force\\
 $R_{rep}$ & 0.75$R_{cc}$ & $l_0$ & Range of repulsive intercellular force\\
 $R_{div}^{0}$ & 0.9 & $l_0$ & Threshold distance for regular cell division\\
 $R_{div}^{l}$ & 1.8 & $l_0$ & Threshold distance for leader cell division\\
  $R_{div}^{ext}$ & 0.9 & $l_0$ & Additional threshold distance for cell division of the graded behaved cells\\
 $k_{div}$ & 0.002 & $1/t_0$ & Division rate for cells surpassing size threshold\\
 $R_{inh}$ & 1.3$l$ & $l_0$ & Contact inhibition range\\
 $\xi$ & 3.0 & $p_0 l_0 t_0$ & Friction coefficient with the substrate for regular cells\\
  $\xi ^ {l}$ & 30.0 & $p_0 l_0 t_0$ & Friction coefficient with the substrate for the leader cell\\
  $\xi ^ {ext}$ & 27.0 & $p_0 l_0 t_0$ & Additional friction coefficient with the substrate for the graded behavior\\
 $t_{relax}$ & 50.0 & $t_0$ & Relaxation time for velocity averaging\\
 $R_{LC}$ & 20 & $l_0$ & Leader cell effects range\\
 $R_{nei}$ & 2.0 & $l_0$ & Boundary cell neighbor range\\
 $D_{max}$ & 0.75 & $l_0$ & Maximum separation for generalized Voronoi construction in Fig 2A\\
 \hline
\end{tabular}

\end{table}